\documentclass[pra,twocolumn,superscriptaddress,amssymb,floatfix,10pt,aps]{revtex4}
\usepackage{graphicx}% Include figure files
\usepackage{algorithm}
\usepackage[noend]{algpseudocode}

\begin{document}

%\tightenlines

\title{Estimating undocumented Covid-19 infections in Cuba by means of a hybrid mechanistic-statistical approach}

\author{Gabriel Gil and Alejandro Lage-Castellanos}
\affiliation{Instituto de Cibern{\'e}tica, Matem{\'a}tica y F{\'i}sica (ICIMAF), La Habana, Cuba, gabriel@icimaf.cu}
\affiliation{Faculta de F{\'i}sica, Universidad de La Habana, La Habana, Cuba, lage@fisica.uh.cu}

\begin{abstract}
We adapt the hybrid mechanistic-statistical approach of Ref.~\cite{roques2020} to estimate the total number of undocumented Covid-19 infections in Cuba. This scheme is based on the maximum likelihood estimation of a SIR-like model parameters for the infected population, assuming that the detection process matches a Bernoulli trial. Our estimations show that (a) $60\%$ of the infections were undocumented, (b) the real epidemics behind the data peaked ten days before the reports suggested, and (c) the reproduction number swiftly vanishes after 80 epidemic days.
\end{abstract}

\maketitle

\section*{INTRODUCTION} % The \section*{} command stops section numbering

%\addcontentsline{toc}{section}{Introduction} % Adds this section to the table of contents

Covid-19 crisis has put forward the need for simple and yet realistic epidemic modeling. The goal is to provide the authorities and the public with accurate predictions to devise and schedule containment and organization policies before an outbreak peaks. To this effect, the daily number of new infections and a minimal description of the epidemic peak (the so-called acme), in terms of the date and number of active infections, are among the crucial data. Another quantity of concern, especially relevant to grasp the full extent of an epidemics --as well as to assess detection strategies-- is the total number of active infections per day. Due to the fact that there is always a fraction of infections that is not detected, \emph{a fortiori} when the pathogen may be carried asymptomatically (e.g., the case of SARS-CoV-2), the full infected population can only be inferred. 

In particular, estimations of total Covid-19 infections (including undocumented trasmission events) are already available in the literature for France \cite{roques2020} and China \cite{li2020}, but are still unknown to many other countries. To the best of our knowledge, there are no current estimates of total Covid-19 infections in Cuba. Therefore, this paper aims at contributing such an important piece of information to the modeling of the pandemics in Cuba.

Historically, epidemic modeling is dominated by mechanistic approaches, like SIS, SIR and SEIR (where letters in the acronyms stands for susceptible, exposed, infected and recovered compartments of the population) \cite{mandal2011}. The advantage of a mechanistic model is that it sets up the epidemic evolution from reasonable time-dependent rules for trasmission and recovery (the mechanics behind the epidemics, so to say). Often, such a mechanical description applies to the full infected population. A straightforward (brute-force) fitting of mechanistic models (e.g., the typical SIR) to just the reported data may not be the correct strategy to follow, since, on one hand, detection draws from statistical sampling of the full infected population, and, on the other, there are detected cases with unknown source of contagion (i.e., unable to stem from the interaction between the infected and susceptible compartments, as modeled).

There have been some efforts directed at reconciling a mechanistic approach with the intrinsic statistical nature of the reported data \cite{breto2009,li2018}. In particular, we are inspired by the hybrid mechanistic-statistical (HMS) approach laid down in Ref.~\cite{roques2020}, which was succesful at modeling Covid-19 in France. 

The HMS scheme by Roques et al.~applies Bayesian inference to estimate SIR parameters (and, hence, the total infected population), assuming that the detection process accomodates to a Bernoulli trial \cite{roques2020}. Since we are interested in limited outbreaks, where government measures are effective at containing the disease, we can choose a simplification of the general SIR model that considers the infected cases negligible with repect to the full population. Moreover, in the same spirit of Cabo-Cabo (fully mechanistic) modeling of Covid-19 in Cuba \cite{cabo2020}, we simulate the effect of the state interventions by means of a heuristic time-dependence of the infection rate, dropping down the day the most strigent measures against spreading were implemented. We also correct the statistical part of the HMS as formulated in Ref.~\cite{roques2020}, by considering that only the still undocumented portion of the infected cases are sampled for test.

The outline of the paper is the following. Sec.~\ref{methods} summarizes the mechanistic and the statistical side of the hybrid approach, emphasizing our ammendments to the formulation in Ref.~\cite{roques2020}. Sec.~\ref{validation} describe the validation scheme and computations. Sec.~\ref{results} tackle two epidemic scenarios: a synthetic outbreak and the case of Covid-19 in Cuba. In Sec.~\ref{conclusion}, we provide some concluding remarks. In the Appendix, we comment on how the synthetic epidemics (used for validation purposes) was generated.

%------------------------------------------------

\section{Methods}\label{methods}

\subsection*{Mechanics: simplified SIR model with a heuristic time-dependent infection rate}\label{SecSIR}

We start from the most customary Susceptible + Infected + Recovered (SIR) model, introduced by Kermack and McKendrick \cite{kermack-mckendrick1927}, i.e.,

\begin{eqnarray}\label{sir}
\frac{dS(t)}{dt} &=& -\alpha \frac{I(t) S(t)}{N} ~,\\
\frac{dI(t)}{dt} &=& \alpha \frac{I(t) S(t)}{N} - \beta I(t) ~,\\
\frac{dR(t)}{dt} &=& \beta I(t) ~.
\end{eqnarray}

\noindent $S(t)$, $I(t)$ and $R(t)$ are the susceptible, infected and recovered time-dependent populations, which sum up to the size of the full population $N$ (i.e., $S(t)+I(t)+R(t)=N$),  whereas $\alpha$ and $\beta$ are the infection and recovery rates. Time $t$ is given in days, hereafter. The cumulative number of infections within a timespan reads

\begin{equation}\label{total}
T(t)=I(t)+R(t) ~.
\end{equation}

At difference with Ref.~\cite{roques2020}, we take a reasonable simplification of the usual SIR model valid for a more or less contained outburst or an early stage of the epidemics \cite{cabo2020}. In such a case, $T(t)\ll N$ and $S(t)\approx N$. Therefore, the infected population reads

\begin{equation}\label{isol}
I(t) = I_0 \exp{\left\{(R_0-1) ~\beta ~(t-t_0)\right\}} ~,
\end{equation}

\noindent where $R_0=\alpha/\beta$ is the basic reproduction number, and $I(t_0)=I_0$ is the initial number of infections. From Eq.~(\ref{isol}), we note that an exponential increase or decrease of infectious events takes place depending on whether $R_0$ is greater or lesser than the unity.

Now, let the infection rate be time-dependent. In such a case, we get

\begin{equation}\label{isolt}
I(t) = I_0 \exp{\left\{\int_{t_0}^t \alpha(t^\prime) ~dt^\prime-\beta ~(t-t_0)\right\}} ~,
\end{equation}

\noindent instead of Eq.~(\ref{isol}). We assume a \emph{heuristic} shape for such a time-dependence, for example, a step function taking a constant value  at the beginning ($\alpha_0>\beta$, to allow for an outbreak) and dropping down to a lower value ($\alpha_{\infty}<\alpha_0$) at some point in time, $t_1$. For a country suffering an epidemics, we set such an inflection point to the day borders were closed or a stringency index jumps abruptly \cite{stringency}. In particular, we choose a Fermi-Dirac distribution as a smooth version of the step function, i.e.,

\begin{equation}\label{alphat}
\alpha(t) = \frac{\Delta\alpha}{1+e^{(t-t_1)/\tau}}+\alpha_{\infty}~,
\end{equation}

\noindent where $\alpha_0\equiv\alpha(t_0)=\Delta\alpha/(1+e^{(t_0-t_1)/\tau})+\alpha_{\infty}$, $\alpha_1\equiv\alpha(t_1)=\Delta\alpha+\alpha_{\infty}$, $\alpha_{\infty}\equiv\lim_{t\to\infty}\alpha(t)$ and $\tau$ is a smooth parameter modulating how fast the infection rate drops down from $\alpha_0$ to $\alpha_{\infty}$. From Eqs.~(\ref{isolt}) and (\ref{alphat}), we get

\begin{eqnarray}\label{isolfinal}
I(t) = &&I_0 \exp{\{(\alpha_1-\beta)(t-t_0)}\} \times\nonumber\\ 
&&\left(\frac{(1 + \exp\{(t_0 - t_1)/\tau\})}{(1 + \exp\{(t - t_1)/\tau\})}\right)^{\Delta\alpha\tau} ~,
\end{eqnarray}

Models relying on heuristic time-dependent infection rates that swiftly vanish after lockdown and stringent measures from the government have been recently explored in Ref.~\cite{cabo2020}.

\subsection*{Statistics: testing as a binomial process and maximum likelihood inference of SIR parameters}\label{SecMLE}

Let us assume that the daily detection of new infected cases $\delta_t$ is a Bernoulli process, as in Ref.~\cite{roques2020}. Hence, the random variable $\delta_t$ distributes binomially,

\begin{equation}\label{dist}
\delta_t \sim \mathrm{Bin}(n_t,p_t) ~,
\end{equation}

\noindent where $n_t$ is the number of trials (i.e., the daily size of the test sample) and $p_t$, the probability of success in each independent trial (for us, the probability of finding infected cases in the population). Subindex $t$ is kept throughout for time series data. 

In such a Bernoulli process, test outcomes are independent from each other. For that to happen, the sample for test would have to be selected at random from the full population. However, more effective and realistic testing strategies often departs from random sampling by focusing on trasmission chains and/or risk groups. To simulate higher prevalence in test samples as compared to random ones, we assume that susceptible population will always be under-sampled, i.e., we bias the probability of finding positive results. In Ref.~\cite{roques2020}, 

\begin{equation}\label{pt}
p_t = \frac{I(t)}{I(t)+\kappa S(t)}
\end{equation}

\noindent where $\kappa\in(0,1)$. 
%We neglect the recovered cases, since those are monitored and kept track of. 

At odds with Eq.~(\ref{pt}), we account for the fact that there is a fraction of the infected population that once tested is quarantined and their retests are not contemplated in $\delta_t$. To that aim, $p_t$ does not builds up from the total but from the currently infected cases that are yet undocumented, i.e., the difference between the full infected population and the active (positive to the test) cases just the day before the current tests, $\mathrm{A}_{t-1}$. Again, we take $S(t)\approx N$. Therefore, Eq.~(\ref{pt}) is replaced by

\begin{equation}\label{ptfinal}
p_t = \frac{I(t)-\mathrm{A}_{t-1}}{I(t)-\mathrm{A}_{t-1}+\kappa N} ~.
\end{equation}

The likelihood of detecting daily reported cases $\delta_t$ given an infected population evolving a l{\`a} SIR (see Eq.~(\ref{isolfinal})) and assuming testing outcomes can be accommodated into a Bernoulli process is \cite{roques2020}

\begin{equation}\label{l}
\mathcal{L}(I_0, \Delta\alpha, \alpha_{\infty}, \tau, \kappa) = \Pi_{t=t_0}^{t_f} \left(\begin{array}{c} n_t \\ \delta_t \end{array} \right) p_t^{\delta_t} (1-p_t)^{n_t-\delta_t} ~.
\end{equation}

\noindent Here, all the parameters are encoded into $p_t$, in the case of SIR parameters ($I_0$, $\Delta\alpha$, $\alpha_{\infty}$ and $\tau$), through the infected population $I(t)$ (see Eqs.~ (\ref{isolfinal}) and (\ref{ptfinal})). Supposing parameters distribute uniformly within the searched domain, the maximization of the likelihood $\mathcal{L}$ lead us to the SIR infected evolution that best reproduce the reported data on daily new cases (in the sense of having the largest probability of being the specific dynamics behind the data). The biasing parameter $\kappa$ is also inferred by maximizing the likelihood.

\section{Validation and computational details}\label{validation}

As anticipated, we want to use our model to estimate the number of total infected population over time from the reported data. Since the motivation for developing such a model is precisely the fact that there are undocumented infections, we do not have an immediate way of validating our results. Roques et al. \cite{roques2020} carried out an indirect validation, by comparing actual data on infected cases with the expected number of the corresponding, binomially distributed, random variable (actually, the cumulative over time of such quantities was employed). Here, we propose as well a direct validation scheme by means of a network epidemic model (NEM). In a nutshell, such a NEM builds up from an unconstrained spread and a detection/quarantine process based on symptomatic cases and traceable trasmission chains they point at (see the Appendix for details). Although, the NEM deserves attention in its own right, its full presentation and analysis is outside the scope of this article and will be published elsewhere. In this paper, it only provides a synthetic epidemic scenario where the total infected population is known by construction, therefore, setting a benchmark for our hybrid model estimation (introduced in earlier sections).

The HMS model is implemented in a Wolfram Mathematica 11.3 notebook (available upon request to the authors). The maximum likelihood estimation (MLE) is carried out through a Nelder-Mead global optimization method with a maximum of $10^3$ iterations. We impose the natural constraints on the parameters (i.e., $\Delta\alpha\ge 0$, $\alpha_{\infty}\ge 0$, $\tau\ge 0$, $I_0>0$ and $\kappa\in(0,1)$) together with $\alpha_0\equiv\Delta\alpha/\left(1+\exp{(t_0-t_1/\tau)}\right)+\alpha_{\infty}>1$, to allow for an initial outbreak. As a matter of fact, we do not maximize the likelihood $\mathcal{L}$ itself (see Eq.~(\ref{l})) but $\ln{(\mathcal{L})}$, which is a smoother function of the parameters. We set $N\approx 11\times 10^6$, approximately the Cuban population. To model the outbreak of Covid-19 in Cuba, we set $1/\beta=20~\mathrm{days}$ \cite{zhou2020} and $t_1$ to the 13th epidemic day, i.e., the day borders and schools were closed and stringency index took the largest leap ($25$ out of a maximum of $100$ units) \cite{data1}. Covid-19 data for Cuba is taken from \cite{data1,data2}. For the MLE in Cuba, we use only the first 80 days, which include an almost complete epidemic peak. Our SIR cannot capture the two-peak profile actually seen in the number of active cases reported for Covid-19 in Cuba (see Fig.~\ref{fig:2}), so far as 110 epidemic days. The number of daily PCR tests for Covid-19 as reported by the Cuban Ministry of Health is a tricky figure in many senses: for example, it is not clear whether it includes or not (a) retests of already detected cases and (b) tests for which the result might be pending \cite{test}. Not having a more intuitive way of estimating the daily number of tests, we choose a constant value for the number of daily tests and set it to the geometric mean of the reported daily Covid-19 tests data in Cuba in a period of 110 epidemic days (i.e., $n_t\approx 1500$). For the synthetic epidemics (our validation case), we change abruptly the reproduction number at the 43rd day (so, we can set $t_1=43~\mathrm{days}$) and choose a recovery time of $1/\beta=7~\mathrm{days}$. The number of tests is the same as for the Covid-19-in-Cuba scenario.

\section{Results}\label{results}

\subsection*{A synthetic epidemic scenario}\label{synthetic}

\begin{figure}[ht!]
    \centering
    \includegraphics[width=0.9\columnwidth]{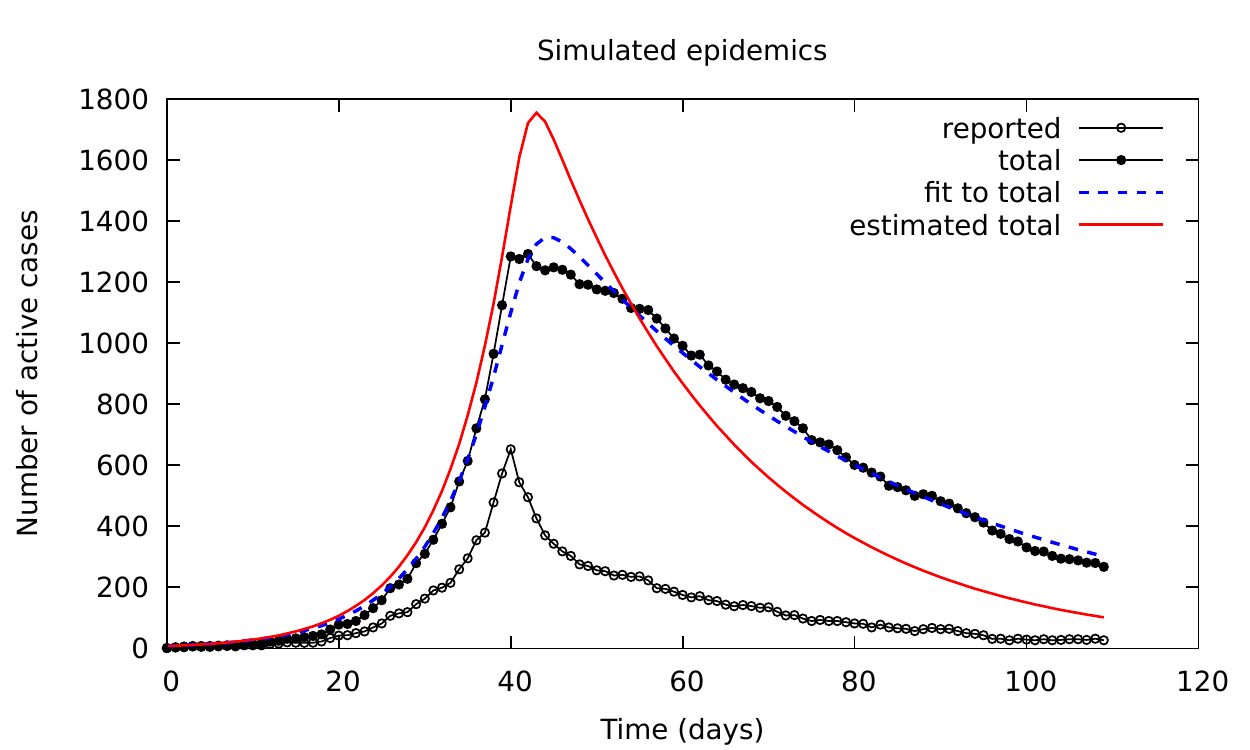}
    \caption{(color online) Total number of active cases for a simulated epidemics (full circles) and our HMS estimation of the same quantity (solid red line). We also show a fit of Eq.~(\ref{isolfinal}) to the total number of active cases (dashed blue line). As a reference, we plot the reported number of active cases (empty circles), which is the input of our HMS estimation.}
    \label{fig:1}
\end{figure}

Figure~\ref{fig:1} shows our results on HMS estimations of the extent of a simulated epidemic outburst, where the total number of active cases is known by construction. Remarkably, such estimations are in a good agreement with the actual benchmark data during the early stages of the outbreak. The quality of our estimation after the epidemics has peaked is poor, but note that they are at least able to spot accurately the date of the acme. Together with our HMS estimations, we plot a non-linear regression model as applied to the total number of active cases. Both the HMS and the regression models have the same functional dependence on time, given by Eq.~(\ref{isolfinal}), and target the same quantity (i.e., the total number of active cases). The practical difference between the two is in the input data: whereas HMS builds up from data on the active and newly detected cases per day, the other directly fits the usually unknown total number of active cases. We do not expect that HMS approach achieve the same degree of success of a fit, since the underlying method does not try to make $I(t)$ conform to the total number of active cases, e.g., by minimizing their relative differences. At odds with a regression, the HMS maximizes the chances of obtaining newly detected cases per day out of a statistical sampling of a proxy for yet undocumented infections (i.e., $I(t)-\mathrm{A}_{t-1}$). The advantage of HMS is that it is useful in all realistic situations in which the total number of active cases is simply not available due to undocumented infections.

\subsection*{Covid-19 in Cuba}\label{cuba}

\begin{figure}[ht!]
    \centering
    \includegraphics[width=0.9\columnwidth]{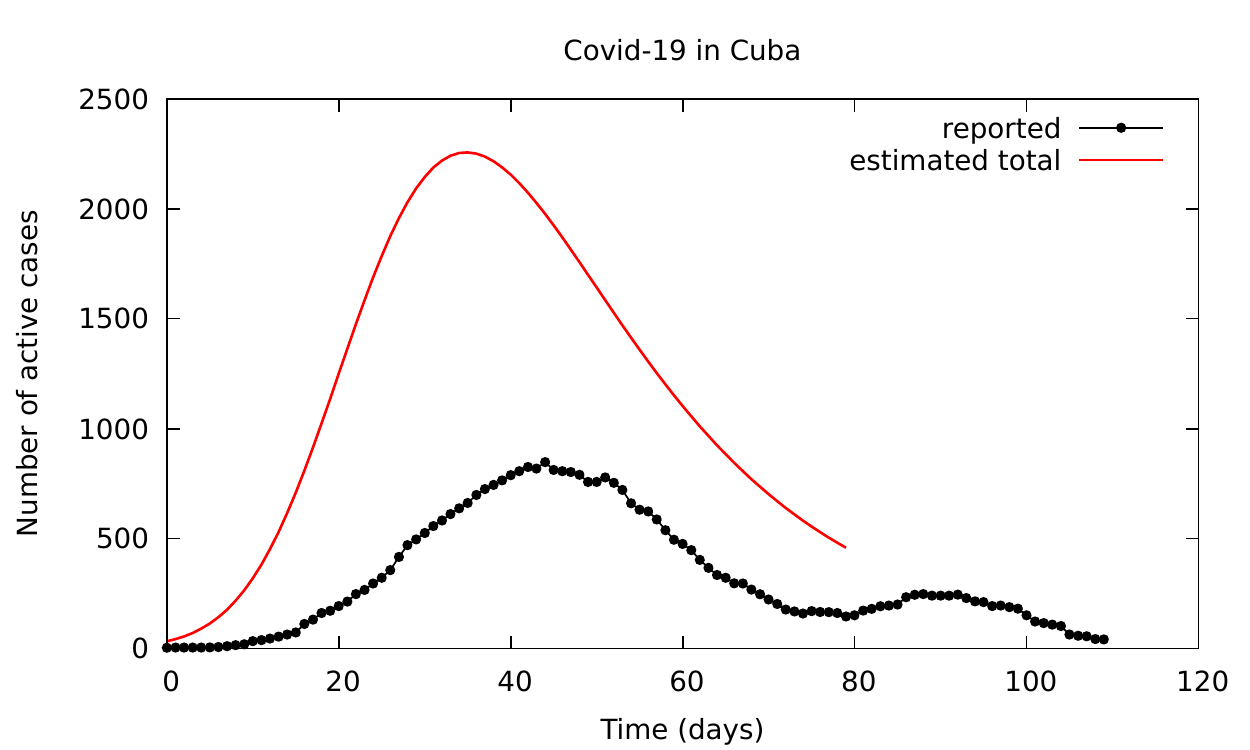}
    \caption{(color online) Number of active cases of Covid-19 in Cuba (black dots), together with our HMS estimation of the total number of active cases (solid red line), including undocumented infections.}
    \label{fig:2}
\end{figure}

As seen in the last section, HMS works well at the epidemic outburst, providing reasonable estimates of the total number of cases and the date of the full epidemic curve (which need not to coincide with the peak of the reported data). Thus validated, we proceed to apply HMS to the recent Covid-19 epidemics in Cuba. In Fig.~\ref{fig:2}, we show a comparison between active cases as obtained from official reports and our estimates (which includes undocumented infections). The full epidemic curve peaks ten days before the reported one. Day by day, reported infections ranges from $4$ to $49\%$ of the total, with a mean of $28\%$, achieved at the full acme. Remarkably, the cumulative number of reported infections within the timespan of 80 days was about $40\%$ of the total (see Eq.~(\ref{total})), and $T/N=4.6\times 10^{-4}$. These are fingerprints of good management of Covid-19 medical crisis in Cuba (cf.~Ref.~\cite{li2020}'s estimate of $14\%$ of documented infections in China before the travel restrictions on 23 January 2020). Moreover, notice that the fraction of documented infections increase over time, from an initial value of $11\%$ to the aforementioned $40\%$, indicating a refinement in Cuban detection process during the epidemics.

In the case of Covid-19 in Cuba, we can only appeal to an indirect validation of our HMS estimates. Fig.~\ref{fig:2} shows the cumulative number of reported infections, $\sum_{t\prime=t_0}^t\delta_{t\prime}$, along with its expected value within our Bernoulli process, $\sum_{t\prime=t_0}^t n_{t\prime}p_{t\prime}$. Good agreement is generally obtained (mean and standard deviation of around 9 and 28 cases, respectively), and nearly zero relative difference at the end of the interval (80 days). 

\begin{figure}[htb!]
    \centering
    \includegraphics[width=0.9\columnwidth]{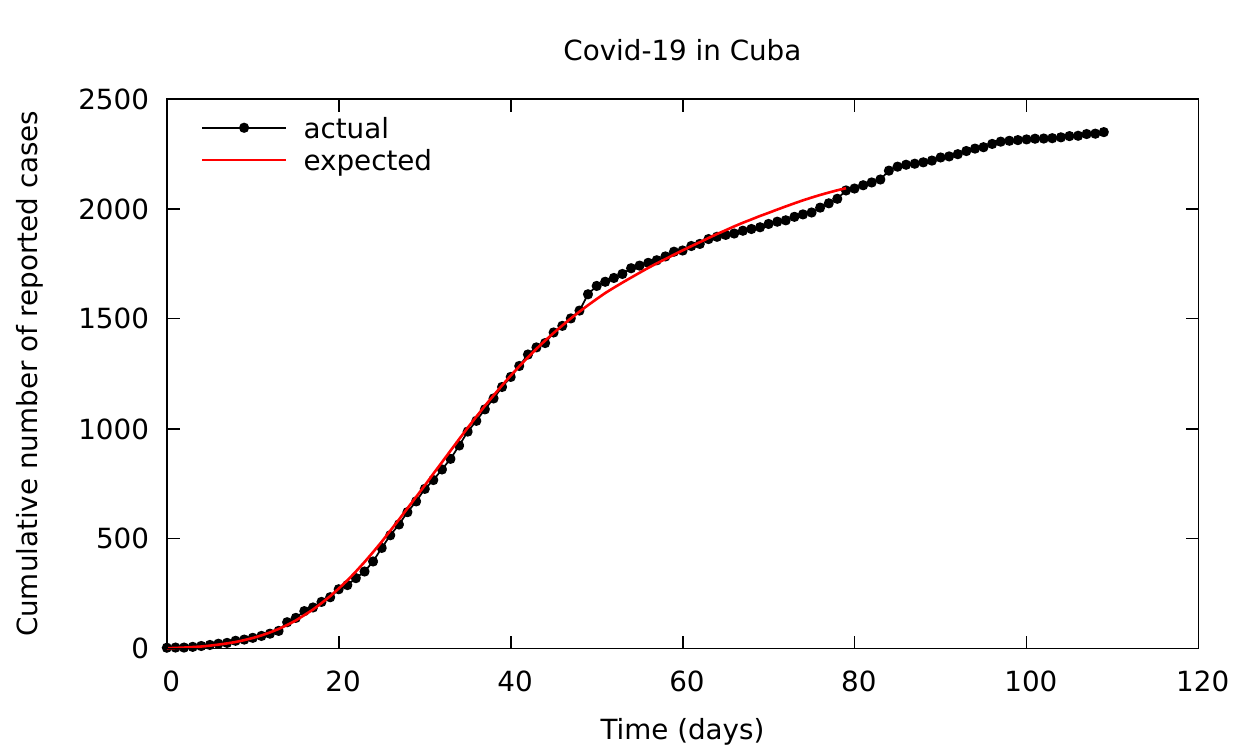}
    \caption{(color online) Cumulative number of reported cases of Covid-19 in Cuba (black dots) along with the expected number of the same quantity by means of HMS (solid red line).}
    \label{fig:3}
\end{figure}

\begin{table}[ht!]
\small
  \caption{Summary of the parameters obtained for each modeling presented in this paper. $R_0^{(0)}=\alpha_0/\beta$ is the reproduction number the first epidemic day, $R_0^{(\infty)}=\alpha_{\infty}/\beta$ is its asymptotic value and $\widetilde{\kappa}=\kappa\times 10^{3}$.}
  \label{tab1}
  \begin{tabular*}{1\linewidth}{@{\extracolsep{\fill}}cccccc}
   \hline
   \hline
  Epidemics & $I_0$ & $R_0^{(0)}$ & $R_0^{(\infty)}$ & $\tau$ (days) & $\widetilde{\kappa}$ \\
   \hline
NEM simulated & 7.81 & 1.92 & 0.69 & 0.86 & 3.9\\  
Covid-19 in Cuba & 32.47 & 6.39 & 0.02 & 11.17 & 4.0\\ 
   \hline
   \hline
   \end{tabular*}
\end{table}

Table~\ref{tab1} summarizes the parameters we get for both, the simulated epidemics and Covid-19 in Cuba. In the latter case, we emphasize the jump in reproduction number the day borders and schools were closed, from 6.39 at the beginning to nearly zero the 80th epidemic day (cf.~\cite{cabo2020}).

\section{Concluding remarks}\label{conclusion}

We adapted the hybrid mechanistic-statistical method developed by Roques et al. \cite{roques2020}, already succesful at modeling Covid-19 in France, to be able to make reasonable estimations of total Covid-19 infections in Cuba. Our theoretical contribution is two-fold. On one hand, we chose a heuristic modification of the classical SIR model that assumes limited outbreaks together with an infection rate changing abruptly when stringent measures take place (see also Ref.~\cite{cabo2020}). On the other hand, we corrected the probability entering the binomial distribution of newly detected cases, in order to account for the fact that the daily reports do not include retest outcomes of already detected cases. Both ammendments turn out to be essential when modeling Covid-19 in Cuba. 

Furthermore, we provided a testing ground for the hybrid mechanistic-statistical estimations: the case of a network epidemic simulation where the total number of active cases is known by construction. The hybrid model is validated against such a benchmark, at least in the early stages of the outburst before the epidemics peaks.

Applying the hybrid model to Covid-19 in Cuba allows us to estimate the total number of active cases, including undocumented infections. The resulting number of undocumented Covid-19 infections in Cuba reaches $60\%$, which is considerably less than the estimate for China ($86\%$) before the travel restrictions were implemented \cite{li2020}, therefore, indicating a good management of the medical crisis in Cuba.

%------------------------------------------------

\section*{ACKNOWLEDGEMENTS} % The \section*{} command stops section numbering
G.G. acknowledges support from the National Programme for Basic Sciences in Cuba, the Cuban Ministry of Science, Technology and Environment (CITMA) and the Abdus Salam International Centre for Theoretical Physics (ICTP) through the grant NT09-OEA. A.L.C. acknowledges PHC Carlos J. Finlay funds, from the Embassy of France in Cuba, for supporting the exchange with french researchers. We were stimulated by the meetings for the mathematical modeling of Covid-19 crisis in Cuba, organized by the Faculty of Mathematics from the University of Havana, in close connection with the Cuban Ministry of Health (MINSAP). In particular, we thank N. Cabo-Bizet and A. Cabo Montes de Oca for sharing and discussing with us an early preprint with their findings, finally included in this RCN special number along with the our paper. G.G. is also grateful to M. Montero and D. S. Fontanella (ICIMAF, Havana, Cuba), for useful discussions and comments.

%------------------------------------------------

\subsection*{Appendix: Network epidemic model with quarantine}

We consider an stochastic branching process in which nodes are infected people, and connections represent the transmission of the disease. As we are interested in the case of controlled or small size epidemics, we will disregard the total size of the population that will be effectively considered as infinite. 

In our simulation, we will sequentially grow an epidemic tree in which nodes are in any of the following states:
\begin{itemize}
 \item[$E$:] exposed to the virus, meaning the person has the virus but is not capable of transmitting it,
 \item[$I_s$:] infectious and symptomatic, meaning the person is capable of transmitting the virus and is also showing symptoms of the disease,
 \item[$I_a$:] infectious and asymptomatic, when the person do not show symptoms but still can transmit the virus,
 \item[$R$:] when the person is no longer transmitting the virus (either because it recovered or died).
\end{itemize}
On top of these states, nodes can either be quarantined or not.

The infectious process is controlled by a set of constants:
\begin{itemize}
 \item[$R_0$:] is the expected number of new infections caused by a single infected individual. This means that in average, every infected person will generate $R_0$ new nodes in the tree;
 \item[$\alpha$:] is the fraction of infected people that will develop symptoms;
 \item[$\beta$:] is the fraction of contacts that are traceable, meaning that if one node is detected to be infected, then it can point to the neighbors (parent or children in the tree) that are connected through traceable contacts;
 \item[$r_{E\to I}$:] is the rate at which exposed nodes turn into infectious;
 \item[$r_{S\to R}$:] is the rate at which infectious and symptomatic nodes recover;
 \item[$r_{A\to R}$:] is the rate at which infectious and asymptomatic nodes recover;
\item[$c_{S}$:] is the rate at which a symptomatic infectious node generates new contacts each day. In order to keep the meaning of $R_0$, we shall have $c_{S} = R_0 \times r_{S\to R}$;
\item[$c_{A}$:] is the rate at which a symptomatic infectious node generates new contacts each day. For the same previous reason, $c_{A} = R_0 \times r_{A\to R}$;
\item[$r_{S\to Q}$:] is the rate at which symptomatic people are detected by the quarantine process and moved to quarantine;
\item[$r_{Q\to R}$:] is the rate at which people are released from the quarantine, either because they died or recovered.
\end{itemize}

\begin{algorithm}
\caption{Stochastic daily SEIRQ cascade process. \label{alg:oneday}}
\begin{algorithmic}[1]
\Procedure{One-day-update}{$E,I_s,I_a,R,Q$ list of nodes in each state}     
    \For{$n \in Q$} 
        \If{$n$ is new in quarantine}
        \State Add-contacts-to-Q($n$) \Comment{Contact tracing}
        \EndIf
        \State Move-Q-to-R-with-prob($r_{Q\to R}$,n)
    \EndFor
    \For{$n \in E$ }
        \State Move-E-to-I-with-prob($r_{E\to I}$,n)
    \EndFor
    \For{$n \in I_s$ }
        \State Generate-offspring-with-rate($c_s$,n)
        \State Move-S-to-R-with-prob($r_{S\to R}$,n)
        \State Move-S-to-Q-with-prob($r_{S\to Q}$,n)
    \EndFor

    \For{$n \in I_a$ }
        \State Generate-offspring-with-rate($c_a$,n)
        \State Move-S-to-R-with-prob($r_{A\to R}$,n)
    \EndFor
\EndProcedure
\end{algorithmic}
\end{algorithm}

The whole simulation is schematized in algorithm \ref{alg:oneday}. Functions Move-A-to-B-with-prob will remove the given node from list A and put it on list B, with a given probability. The function Generate-offspring-with-rate will add new nodes as children of the given node, some of which will be traceable some who won't, and some of which will be symptomatic and some who won't. All this new nodes are added also to the Exposed list. The function Add-contacts-to-Q will follow all the traceable contacts of the node $n$ and put them in quarantine (removing them from the lists they were).

A simulation like this can mimic most of the indicators that are being reported by the Cuban Ministry of Health in its daily briefings. The reports correspond to the characteristics of the nodes that are quarantined each day: whether they come from known contacts, whether they have symptoms or not. A quantity that is not directly included in this simulation is the amount of declared contacts that will be negative to the virus test, since we only deal with positive cases. However, this number is not reported either by the health authorities, and it is natural to assume that the number is a Poisson random variable with not particular other implications in the process.

% \addcontentsline{toc}{section}{Acknowledgments} % Adds this section to the table of contents

%----------------------------------------------------------------------------------------
%	REFERENCE LIST
%----------------------------------------------------------------------------------------

%\bibliographystyle{amsplain}
%\bibliographystyle{babplain}
%\bibliographystyle{plain}
\bibliographystyle{unsrt}
\bibliography{covid}

\end{document}